\documentclass[usenatbib,useAMS,a4paper]{mn2e}
\bibpunct{(}{)}{;}{a}{}{,}

\usepackage[dvips]{graphicx}
\usepackage{times}
\usepackage{color}
\usepackage{appendix}
\usepackage{tabularx}
\usepackage{array}
\usepackage{multirow}
\usepackage{amssymb}

\renewcommand{\theequation}{\thesection.\arabic{equation}}

\newcommand{\hii}{\mbox{H\,{\sc ii}}}
\newcommand{\tauv}{\hbox{$\hat{\tau}_{V}$}}

\newcommand{\taul}{\hbox{$\hat{\tau}_\lambda$}}

\newcommand{\ha}{\hbox{H$\alpha$}}
\newcommand{\hb}{\hbox{H$\beta$}}

\newcommand{\msun}{\hbox{$M_\odot$}}

\newcommand{\oii}{\hbox{[O\,{\sc ii}]}}
\newcommand{\oiii}{\hbox{[O\,{\sc iii}]}}
\newcommand{\nii}{\hbox{[N\,{\sc ii}]}}
\newcommand{\sii}{\hbox{[S\,{\sc ii}]}}

\newcommand{\aboh}{\hbox{$12+\log\textrm{(O/H)}$}}

\newcolumntype{x}[1]{%
{\centering\hspace{0pt}}p{#1}}%

\title[Appropriate spectral models to derive galaxy properties]
{On the importance of using appropriate spectral models to derive physical properties of galaxies at $0.7<z<2.8$}

\author[C. Pacifici, E. da Cunha, S. Charlot et al.]
{Camilla~Pacifici,$^{1,3}$\thanks{E-mail: camilla.pacifici@galaxy.yonsei.ac.kr} Elisabete~da~Cunha,$^{2}$ St\'ephane Charlot,$^{3}$ Hans-Walter Rix,$^{2}$ \and Mattia Fumagalli,$^{4}$ Arjen van der Wel,$^{2}$ Marijn Franx,$^{4}$ Michael V. Maseda,$^{2}$ \and Pieter G. van Dokkum,$^{5}$ Gabriel B. Brammer,$^{6}$ Ivelina Momcheva,$^{5}$ Rosalind E. Skelton,$^{7}$ \and Katherine Whitaker,$^{9}$ Joel Leja,$^{5}$ Britt Lundgren,$^{8}$ Susan A. Kassin,$^{6}$ Sukyoung K. Yi$^{10}$
\\ \\
$^{1}$Yonsei University Observatory, Yonsei University, Seoul 120-749, Republic of Korea\\
$^{2}$Max Planck Institute for Astronomy, K\"{o}nigstuhl 17, D-69117 Heidelberg, Germany\\
$^{3}$UPMC-CNRS, UMR 7095, Institut d'Astrophysique de Paris, 75014, Paris, France\\
$^{4}$Leiden Observatory, Leiden University, P.O. Box 9513, 2300 RA Leiden, The Netherlands\\
$^{5}$Department of Astronomy, Yale University, 260 Whitney Avenue, New Haven, CT 06511, USA \\
$^{6}$Space Telescope Science Institute, 3700 San Martin Drive, Baltimore, MD 21218, USA \\
$^{7}$South African Astronomical Observatory, PO Box 9, Observatory, Cape Town, 7935, South Africa \\
$^{8}$Department of Astronomy, University of Wisconsin-Madison, 475 N. Charter Street, Madison, WI 53706, USA \\
$^{9}$Astrophysics Science Division, Goddard Space Center, Greenbelt, MD 20771, USA \\
$^{10}$Department of Astronomy and Yonsei University Observatory, Yonsei University, Seoul 120-749, Republic of Korea\\
}

\begin{document}

\maketitle

\begin{abstract}
Interpreting observations of distant galaxies in terms of constraints on physical parameters -- such as stellar mass ($M_{\ast}$), star-formation rate (SFR) and dust optical depth ($\tauv$) -- requires spectral synthesis modelling. We analyse the reliability of these physical parameters as determined under commonly adopted `classical' assumptions: star-formation histories assumed to be exponentially declining functions of time, a simple dust law and no emission-line contribution. Improved modelling techniques and data quality now allow us to use a more sophisticated approach, including realistic star-formation histories, combined with modern prescriptions for dust attenuation and nebular emission \citep{pacifici2012}. We present a Bayesian analysis of the spectra and multi-wavelength photometry of 1048 galaxies from the 3D-HST survey in the redshift range $0.7<z<2.8$ and in the stellar mass range $9\lesssim\log(M_{\ast}/M_{\odot})\lesssim12$. We find that, using the classical spectral library, stellar masses are systematically overestimated ($\sim 0.1$\,dex) and SFRs are systematically underestimated ($\sim 0.6\,$dex) relative to our more sophisticated approach. We also find that the simultaneous fit of photometric fluxes and emission-line equivalent widths helps break a degeneracy between SFR and $\tauv$, reducing the uncertainties on these parameters. Finally, we show how the biases of classical approaches can affect the correlation between $M_{\ast}$ and SFR for star-forming galaxies (the `Star-Formation Main Sequence'). We conclude that the normalization, slope and scatter of this relation strongly depend on the adopted approach and demonstrate that the classical, oversimplified approach cannot recover the true distribution of $M_{\ast}$ and SFR.
\end{abstract}


\section{Introduction}

Our understanding of galaxy evolution has immensely progressed over the last decades thanks to deep photometric observations mainly at ultraviolet, optical and near-infrared wavelengths enabled by the {\it Hubble Space Telescope} (HST). The {\it Spitzer} and {\it Herschel} space telescopes have complemented such datasets at mid- and far-infrared wavelengths, allowing us to characterize galaxy populations out to high redshifts in unprecedented detail.
These observations suggest that the Universe has undergone a peak of star formation activity at $1<z<3$ (e.g.~\citealt{hopkins2006,bouwens2007,cucciati2012}). Mapping this epoch of intense galaxy assembly is thus crucial to understand galaxy evolution.
The new WFC3 camera on board HST has revolutionized the study of the galaxy population at this epoch. In particular, the Cosmic Assembly Near-IR Deep Extragalactic Legacy Survey (CANDELS; \citealt{grogin2011,koekemoer2011}) HST Multi-Cycle Treasury Survey is providing high spatial resolution images of the emission in the most extensively studied extragalactic fields (GOODS, COSMOS, EGS and UDS) for which a wealth of ancillary photometric data from the ultraviolet to the infrared have been compiled over the years. At the same time, the 3D-HST WFC3-G141 grism survey has been obtaining spatially resolved rest-frame optical spectra of $\sim 7000$ galaxies ($H < 23$) at $z>1$ in these fields, providing redshift estimates with an accuracy of $\sigma(z)=0.0034(1+z)$, i.e. an order of magnitude better than that of photometric redshift based on broad-band observations \citep{brammer2012}. Crucially, these new spectroscopic data also enable detailed studies of the ionized gas and dust geometry (via rest-frame optical emission lines) and stellar population properties (via stellar absorption features) of unprecedentedly large galaxy samples at $z>1$ (e.g.~\citealt{brammer2012,fumagalli2012,fumagalli2013,schmidt2013,whitaker2013,price2014}).

A main caveat in current statistical studies of the galaxy population at $z\ga1$ is that the way in which the physical properties of galaxies are generally derived from rich multi-wavelength data-sets does not reflect recent advances in the sophisticated modelling of galaxy spectral energy distributions (SEDs). For example, spectral analyses often rely on oversimplified colour diagnostics and/or limited modelling of the stellar spectral continuum using simple star formation histories (SFHs), such as exponentially declining  $\tau$-models.\footnote{SFHs in which the star formation rate depends on time $t$ as $\propto\exp(-t/\tau)$ are often referred to as `$\tau$-models'.} As shown for example by \cite{maraston2010}, such models do not account well for the observed colours of $z\approx 2$ galaxies from the GOODS-South sample, which these authors reproduce by appealing to exponentially rising SFHs \cite[see also][]{pforr2012}.  \cite{simha2014} propose a 4-parameter model (delayed exponential + linear ramp at some transition time) to best represent the SFHs of galaxies from smoothed particle hydrodynamics simulations. Other studies have already shown that more sophisticated SFH parametrizations provide better agreement with the data (e.g.~\citealt{lee2010,pacifici2013,behroozi2013,lee2014}). The inclusion of nebular emission is also important to interpret observed spectral energy distributions of galaxies. While most studies including nebular emission rely on empirically calibrated emission-line template spectra (e.g.~\citealt{fioc1997,anders2003,schaerer2009,schaerer2010}), more elaborate prescriptions have been proposed, based on combination of stellar population synthesis and photoionization codes (e.g., \citealt{charlot2001,groves2008}).

\cite{pacifici2012} provide the sophisticated modelling framework required to interpret multi-wavelength photometric and spectroscopic galaxy observations in a physically and statistically consistent way. This approach overcomes several main limitations of galaxy spectral modelling mentioned above. Specifically, \cite{pacifici2012} build a comprehensive library of model galaxy SEDs, which can be used to derive statistical constraints on physical parameters. This combines: (i) physically motivated star formation and chemical enrichment histories from cosmological simulations; (ii) state-of-the-art stellar population synthesis and nebular emission modelling computed consistently using the photoionization code {\small CLOUDY} \citep{ferland1996} and (iii) a sophisticated treatment of dust attenuation, which includes uncertainties in the spatial distribution of dust and in galaxy orientation \citep{chevallard2013}. One of the main features of the \cite{pacifici2012} approach is that it allows one to interpret simultaneously the stellar and nebular emission from galaxies at any spectral resolution.

In this paper, we use the wealth of multi-wavelength data provided by the 3D-HST survey to investigate, in a systematic way, how different SED modelling approaches (regarding star formation and chemical enrichment histories, dust attenuation and nebular emission lines) and the availability of spectroscopic (in addition to photometric) information affect the constraints derived on the physical parameters of high-redshift galaxies. We use the \cite{pacifici2012} models to reproduce simultaneously the observed $0.35$ to $3.6\mu$m photometry and emission line strengths of a sample of 1048 galaxies at $0.7<z<2.8$. We investigate the advantages provided by these sophisticated models when deriving galaxy physical parameters, with respect to standard approaches relying on widely used, oversimplified parametrizations of star formation histories (continuous, exponentially declining $\tau$-models), fixed assumptions about metallicity and dust properties and neglecting nebular emission (e.g.~\citealt{noll2009,kriek2009,forster2009,marchesini2009,pozzetti2010}). We also quantify the importance of accounting for nebular emission in the interpretation of broad-band galaxy observations by comparing the results obtained using the original \cite{pacifici2012} models with those obtained using the same sophisticated prescriptions for the star formation and chemical enrichment histories and attenuation by dust, but not including nebular emission lines.
To further illustrate the importance of using appropriate spectral models to derive physical properties of galaxies and, more generally, to constrain galaxy evolution, we also explore how the different SED-fitting methods considered in this paper affect the interpretation of the correlation between SFR and stellar mass of galaxies (commonly referred to as the `star-formation main sequence' - \citealt{noeske2007}) and the scatter about this relation.

Our paper is organized as follows. In Section~\ref{sec:data}, we describe the sample of 1048 galaxies in GOODS-South extracted from the 3D-HST catalogue, for which multi-wavelength photometry and HST/WFC3 grism spectra are available. In Section~\ref{sec:modelintro}, we introduce three model libraries of galaxy SEDs with different levels of sophistication to interpret these observations. We also assess how well each of these libraries can account for the data. Then, in Section~\ref{sec:results}, we quantify the differences in the constraints derived on the stellar masses, SFRs and attenuation by dust for the galaxies in the sample when adopting the different spectral libraries. We investigate in Section~\ref{sec:ms} how these differences affect the measured correlation between SFR and stellar mass of galaxies. Finally, in Section~\ref{sec:summary}, we summarize our results. Throughout this paper we express magnitudes in the AB system and we adopt the \cite{chabrier2003} stellar IMF and the following cosmology: $\Omega_\Lambda=0.7$, $\Omega_m=0.3$, $H_0=70$ km s$^{-1}$ Mpc$^{-1}$.

\section{Data}
\label{sec:data}

We select a sample of galaxies from the 3D-HST Treasury Survey\footnote{http://3dhst.research.yale.edu/} \citep[PI: van Dokkum]{brammer2012}, which provides a multi-wavelength photometric catalogue \citep{skelton2014} and low-resolution WFC3 grism observations. In this Section, we present the photometric catalogue, spectroscopic observations and describe the sample selection.

\subsection{Photometry}
\label{sec:phot}

We use version 4.1 of the 3D-HST Survey photometric catalogue for the GOODS-South field covering an area of 171 arcmin$^2$ \citep{skelton2014}. The catalogue contains broad- and narrow-band photometry in 38 filters sampling ultraviolet (UV) to mid-infrared (MIR) wavelengths, from both ground- and space-based observatories, for a total of 50,507 objects detected in the HST/WFC3-F160W band. To make the present investigation relevant to a wide range of studies of high-redshift galaxies, we focus on the following commonly used nine broad bands: $U$, ACS-F435W, ACS-F606W, ACS-F775W, ACS-F850lp, WFC3-F125W, WFC3-F140W, WFC3-F160W and {\em Spitzer}/IRAC 3.6$\mu$m. These bands allow us to reasonably sample the rest-frame UV, optical and near-infrared emission of galaxies at all redshifts between 0.7 and 2.8, optimally probing the emission by stellar populations of different ages and the effects of dust attenuation. We exclude {\em Spitzer}/IRAC data at wavelengths longer than $3.6\mu$m because the spectral modelling (see Section~\ref{sec:modelintro}) does not include dust emission.\footnote{Adding dust emission increases the complexity of the analysis and is beyond the scope of the current paper; we leave a detailed analysis combining the \citealt{pacifici2012} model with the models by \citealt{dacunha2008} to a future study.}

The photometry in all bands is extracted consistently across all wavelengths as described in \cite{skelton2014}. To summarize, the 3D-HST photometric catalogue is based on WFC3 selection using a combination of the three WFC3 bands (F125W, F140W and F160W) for detection. In order to measure consistent colours across different HST bands, each of the HST images is convolved with the same point spread function (PSF) and aperture photometry is performed on the PSF-matched images. For the lower resolution ground-based optical and {\em Spitzer}/IRAC bands, the photometry is extracted using the method described in \cite{labbe2006}, \cite{wuyts2008} and \cite{whitaker2011}, which uses the HST images as a high-resolution prior to correct for the larger PSF and confusion caused by neighboring sources.

\subsection{Grism spectroscopy}
\label{sec:spec}

The 248 orbits dedicated to the 3D-HST Survey are distributed among 124 individual pointings, each observed for two orbits. 34 of these pointings cover the GOODS-South field. For all 50,507 galaxies in this field, WFC3-G141 grism spectra are available, while a spectroscopic redshift ($z_\mathrm{grism}$) has been estimated for a subset of 2699 galaxies to $H<23$ \citep[here the $H$ band is HST/WFC3-F160W;][]{brammer2012}.
The WFC3 grism spectra cover observer-range wavelengths between 1.1 and 1.6 $\mu$m. The spectral scale is $\approx 23$ {\AA}/pixel and the effective resolution depends on the spatial extent of the galaxy (see \citealt{brammer2012}, Sec. 2.1 and 4.2). For example, the observed full width at half maximum (FWHM) of the {\ha} line (which in these spectra is blended with $\nii\lambda 6548$ and $\nii\lambda 6584$) can span a range as wide as 75--200 \AA. The emission-line sensitivity is $\sim 5 \times 10^{-17}$ erg s$^{-1}$ cm$^{-2}$ at $5\sigma$. 
Grism redshifts are derived from a combination of the 1D spectra and broad- and medium-band photometry from the catalogue described above (see \citealt{brammer2012} and references therein for details). The precision of grism redshifts is $\sigma(z)=0.0034(1+z)$, i.e. an order of magnitude better than what is typically achieved with high-quality broad-band photometry alone.

To characterize the emission features in the observed spectra, we measure rest-frame equivalent widths (EWs) of the most prominent lines. Because of the low resolution, some emission lines are blended together, thus we extract the EWs of: ${\oii}\lambda\lambda3726,\,3729$; {\hb}; ${\oiii}\lambda\lambda4959,\,5007$; $\nii\lambda6548+\ha+\nii\lambda6584$; and ${\sii}\lambda\lambda6717,\,6731$. In the following, we refer to these (blends of) lines simply as {\oii}, {\hb}, {\oiii}, {\ha} and {\sii}. To measure the continuum underneath the emission lines, we first smooth the spectra over 17 pixels (i.e., about 200 \AA\ in rest-frame at $z=1$). Then, we mask $\approx350${\AA}-wide regions centred on the 5 (groups of) emission lines, interpolate the spectra in the masked regions and finally smooth again over 17 pixels to remove possibly persisting features. Using the resulting continuum, we can calculate the emission-line EWs of {\oii}, {\hb}, {\oiii}, {\ha} and {\sii} as defined above. The uncertainties on the EWs are measured through the propagation of the errors on the flux.\footnote{Since the continuum is obtained by smoothing the flux, we scale the error on the continuum with the fraction between the standard deviations of the flux and the continuum in a wavelength window free of line contamination.} The signal-to-noise ratio (S/N) of {\ha} and {\oiii} can reach up to $\sim30$, while that of the other lines ({\oii}, {\hb} and {\sii}) rarely exceeds $\sim10$.

\begin{figure}
\begin{center}
\includegraphics[width=0.5\textwidth]{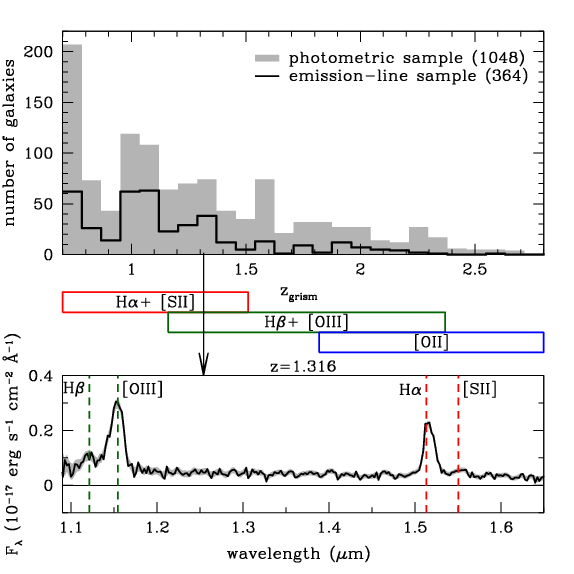}
\caption{Redshift distributions of the 3D-HST photometric sample (gray shaded histogram; 1048 galaxies) and emission-line sub-sample (black open histogram; 364 galaxies) are shown in the top panel. Coloured boxes indicate the redshift windows in which selected emission lines can be detected in the WFC3-G141 grism spectra. In the bottom panel, we show the observer-frame grism spectrum of one of the highest S/N objects in the sample, detected at $z=1.316$. The black solid line and the grey shaded region represent the observed flux and the flux error, respectively. Dashed lines indicate the position of the four detected (blends of) emission lines.}
\label{fig:zz}
\end{center}
\end{figure}

\subsection{Sample selection}
\label{sec:sample}

\begin{table}
\begin{center}
\caption{Typical magnitudes and signal-to-noise ratios of the 3D-HST photometric sample in the nine bands considered in this paper.}
\label{tab:phot}
\begin{tabular}{l r r r}
\hline
\hline
Filter & Effective $\lambda$/{\AA} & Median magnitude & Median S/N \\
\hline
$U$ & 3446 & 25.1 & 52\\
ACS-F435W & 4302 & 24.9 & 23\\
ACS-F606W & 5925 & 24.3 & 52\\
ACS-F775W & 7705 & 23.6 & 62\\
ACS-F850lp & 9059 & 23.2 & 72\\
WFC3-F125W & 12,516 & 22.5 & 91\\
WFC3-F140W & 13,971 & 22.4 & 61\\
WFC3-F160W & 15,392 & 22.2 & 105\\
IRAC 3.6$\mu$m & 35,573 & 21.5 & 106\\
\hline
\hline
\end{tabular}
\end{center}
\end{table}

We first select a \textit{photometric} sample by extracting from the 3D-HST v4.1 photometric catalogue all galaxies with a measured grism redshift in the range $0.7<z_\mathrm{grism}<2.8$ and use\_flag $=1$ (see Section 3.8 in \citealt{skelton2014}). This sample includes 1140 galaxies down to $H=23$. From this sample, we select an \textit{emission-line} sub-sample including only galaxies with at least one emission line detected in the WFC3-G141 grism spectrum with equivalent-width S/N greater than 5. We visually inspect all selected grism spectra to check the accuracy in redshift and to make sure that contiguous emission lines ({\hb} with {\oiii} and {\ha} with {\sii}) are well separated. In the end, we retain 392 emission-line galaxies. We then reject all sources detected by \textit{Chandra} (\citealt{xue2011}, $\log(L_\mathrm{X})>42$) as potential AGNs, which leaves us with 1048 galaxies in the photometric sample and 364 in the emission-line sample. In Table~\ref{tab:phot}, we summarize the median magnitudes and S/N in the nine filters of interest to us for the photometric sample. We note that the typical S/N on the photometry is very high for most bands. We discuss the influence of such small uncertainties on the derivation of statistical constraints on physical parameters in Section~\ref{sec:fit} and \ref{sec:results}.

The redshift distributions of the photometric and emission-line samples are shown in Fig.~\ref{fig:zz} (top panel) as shaded and open histograms, respectively. The peak at $z\approx0.7$ is associated to an over-density in the field (\citealt{salimbeni2009}). Most of the galaxies in the emission-line sample lie at $z\sim1$, which corresponds to where {\ha} falls in the G141 spectral range. The most `data-rich' window is the redshift range $1.2 \lesssim z \lesssim 1.5$, where four emission lines can be detected simultaneously; 17 per cent of the galaxies in the emission-line sample are selected from this redshift window. As an example, we show, in the bottom panel of Fig.~\ref{fig:zz}, the observer-frame grism spectrum of one of the highest S/N objects in the sample, detected at $z=1.316$. This galaxy shows two emission lines with S/N~$>20$ ({\ha} and {\oiii}), one with S/N~$\approx10$ ({\hb}) and one with S/N~$\approx5$ ({\sii}).

\section{Comparison between different spectral modelling approaches}
\label{sec:modelintro}

To derive galaxy physical parameters (such as stellar mass, SFR and optical depth of the dust) from the multi-wavelength observations described above, we must appeal to spectral modelling techniques. A main focus of the present paper is to establish the appropriateness of different spectral modelling approaches to interpret photometric and spectroscopic observations of distant galaxies. Specifically, we consider three modelling approaches relying on different assumptions about several main ingredients of spectral interpretation techniques: the explored (prior) ranges of star formation and chemical enrichment histories; attenuation by dust; and nebular emission. In the next paragraphs, we describe these competing approaches in order of increasing complexity. We also quantify the extent to which the different models can account for the data of Section~\ref{sec:data} by comparing prior libraries of predicted colours with these multi-wavelength data with a bayesian approach. We refer to Appendix A for a specific comparison between physical parameters extracted using our most sophisticated spectral library (P12, see below) and a tool widely used in the 3D-HST collaboration \citep[FAST]{kriek2009}.

\subsection{Different model spectral libraries}
\label{sec:model}

\subsubsection{The `classical' spectral library (CLSC)}
\label{sec:basic}

We first assemble a `classical' model spectral library, corresponding to standard simplistic descriptions of the stellar and interstellar content of galaxies, which are the most widely used to derive galaxy physical properties from fits of ultraviolet to near-infrared SEDs (e.g.~\citealt{kriek2009,pozzetti2010}). Galaxy SFHs in this spectral library are parametrized as exponentially declining functions of the form $\psi(t)\propto \exp(\gamma \, t)$. Here $\gamma$ is the inverse star formation timescale, drawn randomly in the range $0<\gamma/\mathrm{Gyr}^{-1}<3$,\footnote{We note that allowing for larger values of $\gamma$ would increase the number of low-specific-SFR galaxies in the prior at fixed evolutionary stage. We have checked that, on the one hand, this would bias the SFR estimates low \citep{wuyts2011,price2014}. On the other hand, very young ages (and thus low masses) would be required in order to match the photometry of the galaxies in the sample. In Appendix A, we show the physical parameters extracted using the code FAST \citep{kriek2009}, in which very short $e$-folding star-formation timescales are allowed.} and $t$ is the lookback time. In this CLSC spectral library, all stars in a given galaxy are assumed to have the same metallicity, drawn randomly from the logarithmic range $-1.6<\log(Z/Z_{\odot})<0.4$ (we adopt the solar metallicity $Z_{\odot}=0.017$). By analogy with \cite{pacifici2012}, we generate a sample of 4 million galaxy SFHs, selecting randomly the redshift of observation in the range $0.6<z<3.0$ (appropriate for the 3D-HST sample described in Section~\ref{sec:data}) and the evolutionary stage between 0.5 Gyr and the age of the Universe at the redshift of observation (see Sections 2.1 and 3.1.2 of \citealt{pacifici2012}).\footnote{In \cite{pacifici2012}, the evolutionary stage is drawn randomly from a linear range in redshift. Here, we choose to draw the evolutionary stage randomly from a linear range in lookback time to avoid having in the spectral library too many galaxies which started forming stars less than 1 Gyr before the redshift of observation.} We combine this set of exponentially declining star formation histories with the latest version of the the \cite{bruzual2003} stellar population synthesis models, which allows one to compute the spectral evolution of stellar populations for given SFH, metallicity and IMF (see Appendix B for a comparison with the original version of the \citealt{bruzual2003} stellar models). We model interstellar dust attenuation using a two-component model, where the $V$-band attenuation optical depth seen by stars older than 10Myr in the diffuse interstellar medium (ISM) is a fraction ($\mu$) of the optical depth $\tauv$ seen by younger stars still embedded in their birth clouds \citep{charlot2000}. We adopt a fixed slope of the attenuation curve in both the birth clouds and the diffuse ISM ($\taul \propto \lambda^n$, with $n=-0.7$). In this spectral library, we draw randomly the total attenuation optical depth $\tauv$ in the range between 0 and 4 and fix $\mu=0.3$. The overall (i.e. galaxy-wide) attenuation law produced by the Charlot \& Fall (2000; see their fig. 5) model depends on the relative contributions by young and old stars to the emission of a galaxy and for young starbursts, can resemble the \cite{calzetti2000} attenuation law (as used in several studies, e.g.~\citealt{kriek2009,forster2009,marchesini2009}).

\subsubsection{The `P12 without emission lines' spectral library (P12nEL)}
\label{sec:p12}

The second model spectral library we consider is based directly on the approach of \cite{pacifici2012}. It includes physically motivated star formation and chemical enrichment histories derived from a post-treatment of the Millennium cosmological simulation \citep{springel2005} using the semi-analytic models of \cite{delucia2007}. This approach allows us to explore a wide range of (non-parametric) SFHs, including declining, rising, roughly constant, bursty and smooth evolutionary shapes, in addition to realistic chemical enrichment histories. As in the case of the CLSC spectral library, we first select randomly the redshift of observation and the evolutionary stage of each model galaxy. Then, following \cite{pacifici2012}, to further widen the range of physical properties probed by this library, and we re-sample the `current' (i.e. averaged over a period of 10Myr before a galaxy is looked at) star formation and chemical properties of each model galaxy by redrawing the specific SFR (the SFR divided by the stellar mass; $\psi_S = $~SFR~$/M_{\ast}$) in the range $-2<\log(\psi_S/\mathrm{Gyr}^{-1})<+2$ and the gas-phase oxygen abundance in the range $7<\aboh<9.4$. For each star formation and chemical enrichment history in the original library we draw 10 different realizations of these current properties and assemble in this way a library of 4 million star formation and chemical enrichment histories. We then generate a library of 4 million model spectra by combining this set of star formation and chemical enrichment histories with the same stellar population synthesis models as used to build the CLSC spectral library in Section~\ref{sec:basic} above. We include dust attenuation using a more sophisticated implementation of the \cite{charlot2000} two-component dust model than in the CLSC spectral library. In this prescription, the slope of the attenuation curve is fixed in the birth clouds ($n=-1.3$), but is drawn randomly in the range $-1.1<n<-0.4$ in the diffuse ISM to reflect uncertainties about the spatial distribution of dust and the orientation of a galaxy (see \citealt{pacifici2012,chevallard2013}). As in the case of the CLSC spectral library, we take $\tauv$ to be randomly distributed between 0 and 4. In summary, the P12nEL spectral library differs from the CLSC spectral library of Section~\ref{sec:basic} in that it includes far more realistic star formation and chemical enrichment histories and a more sophisticated prescription for attenuation by dust.

\subsubsection{The `P12 with emission lines' spectral library (P12)}
\label{sec:p12el}

Finally, the third model spectral library we consider is the original one of \cite{pacifici2012}. This is identical to the P12nEL library described in the previous section, except that it also includes a component of nebular emission. \cite{pacifici2012} used the standard photoionization code {\small CLOUDY} \citep{ferland1996} to compute the emission lines consistently with the emission by stellar populations of different ages (neglecting the contribution by stars older than 10 Myr, which produce negligible ionizing radiation). The nebular emission in a model spectrum is parametrized in terms of the gas-phase metallicity, $Z$, the zero-age ionization parameter, $U_0$ and the dust-to-metal (mass) ratio, $\xi_d$ (which characterizes the depletion of metals onto dust grains) of the photoionized gas (see section 2.2.2 of \citealt{pacifici2012} for more details). These parameters are randomly drawn while re-sampling of the current chemical properties of a galaxy (Section~\ref{sec:p12}). We build in this way a P12 library of 4 million model galaxy spectra.

\subsection{Confronting spectral models with the observations}
\label{sec:mockobs}

\begin{figure*}
\begin{center}
\includegraphics[width=0.8\textwidth]{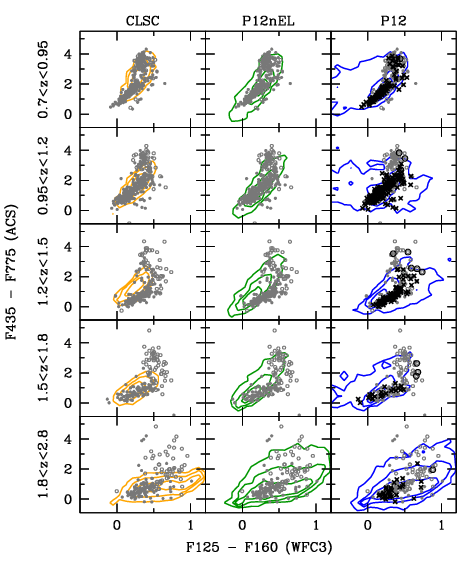}
\caption{Optical-NIR colour-colour diagrams comparing the two 3D-HST samples with the three model libraries. The photometric sample and the emission-line sub-sample (including only galaxies with at least one emission line detected with equivalent-width S/N~$>5$) are represented by gray and black symbols, respectively; open circles mark objects for which the error in at least one of the two colours is larger than 0.2 magnitudes. Contours show the colour-colour space covered by the three spectral libraries: CLSC (left-hand column, orange), built using exponentially declining SFHs, fixed metallicity, simple dust attenuation and no nebular emission (Section~\ref{sec:basic}); P12nEL (middle column, green), built using physically motivated star formation and chemical enrichment histories and a sophisticated treatment for dust attenuation (Section~\ref{sec:p12}); P12 (right-hand column, blue), same as previous, including a component of nebular emission (Section~\ref{sec:p12el}). In each panel, the three contours mark 50, 16 and 2 per cent of the maximum density. While the CLSC spectral library leaves few observed galaxies with no model counterpart, the P12 spectral library allows us to cover reasonably well the entire observed colour-colour space.}
\label{fig:col3lib}
\end{center}
\end{figure*}

Hence, we have built three model spectral libraries relying on different prescriptions to describe galaxy star formation and chemical enrichment histories, attenuation by dust and nebular emission. Each spectral library contains 4 million galaxy SEDs covering the rest-frame wavelength range $912\,\mathrm{\AA}<\lambda<5\,\mu \mathrm{m}$.\footnote{We do not include here reradiation by dust grains, which is expected to dominate the emission at wavelengths greater than $\sim 5\,\mu$m (e.g., \citealt{dacunha2008}).} To compare these model SEDs with the observations of galaxies in the 3D-HST photometric and emission-line samples, we convolve the observer-frame model SEDs with the response functions of the filters presented in Section~\ref{sec:phot}. For P12, we also compute the EW of the emission lines in each model spectrum in the same way as we compute the observed EWs from the WFC3 grism spectra. In brief, we first reproduce the average resolution of the observations ($100$ {\AA} FWHM and 22.5{\AA} bin width) and then calculate the EWs of the {\oii}, {\hb}, {\oiii}, {\ha} and {\sii} emission lines (as defined in Section~\ref{sec:spec}).

In Fig.~\ref{fig:col3lib}, we compare the observations of both the photometric and emission-line 3D-HST samples described in Section~\ref{sec:sample}, in two ACS and two WFC3 bands, with the predictions of the three model spectral libraries.  We plot the ACS (F435W $-$ F775W) vs WFC3 (F125W $-$ F160W) observer-frame colours in different redshift ranges (rows) for the three different spectral libraries (columns). Contours show the colour-colour space covered by the different spectral libraries. The photometric sample (grey symbols) and the emission-line sample (black symbols) are shown on top. Since the emission-line sample can be analysed only with the P12 spectral library, we plot it only in the right hand-side column. This figure shows that the CLSC spectral library leaves few observed galaxies with no model counterpart. Thus, SED fits for these galaxies will be biased towards the models that lie the closest to the observations, at the very edge of the spectral library. The P12nEL spectral library, which does not include nebular emission, can cover reasonably well the bulk of the observations at all redshifts. This shows the importance of accounting for more realistic ranges of star formation (and chemical enrichment) histories and dust properties than included in the widely used CLSC spectral library. Few observed galaxies fall outside the contours of the P12nEL model spectral library, presumably because of the contamination of the WFC3-F160W flux by strong {\ha} emission. In fact, the P12 spectral library, which includes contamination of broad-band fluxes by nebular emission, spans a much larger range of the colour-colour space than the CLSC and P12nEL libraries in Fig.~\ref{fig:col3lib}. This spectral library allows us to cover reasonably well the entire observed colour-colour space, with the exception of some fairly red galaxies with large photometric errors (empty circles).

We note that, the emission-line 3D-HST sample in Fig.~\ref{fig:col3lib} (right-hand panel), although including only galaxies with at least one well-detected emission line, is not biased towards the strongest starburst galaxies (characterized by blue ACS colours), but populates a similar colour-colour space as the photometric sample. This is because the selection in equivalent-width S/N also gathers those massive galaxies with good continuum S/N and relatively faint emission lines. The global properties of the photometric and emission-line 3D-HST samples in Fig.~\ref{fig:col3lib} are thus quite similar.

\subsection{Spectral fits}
\label{sec:fit}

\begin{figure*}
\begin{center}
\includegraphics[width=0.8\textwidth]{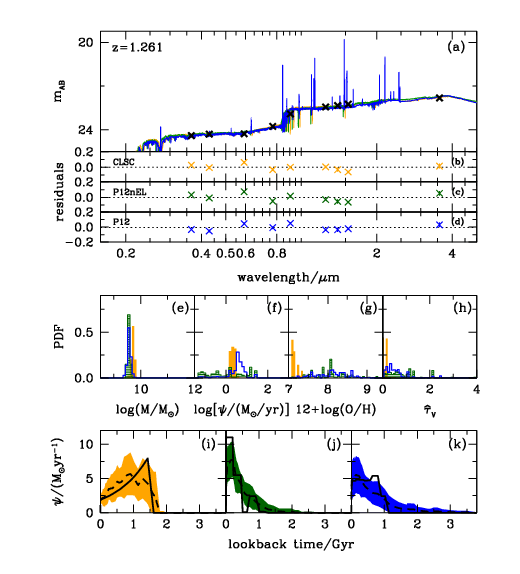}
\vspace{-0.3cm}
\caption{An example of spectral fit of a galaxy at redshift $z=1.261$ using the three model spectral libraries. (a) Observed broad-band magnitudes (black crosses) and best-fitting observer-frame model spectra in full resolution computed with the CLSC (orange), P12nEL (green) and full P12 (blue) spectral libraries. (b, c, d) Residuals between the observed magnitudes and the magnitudes of the best-fitting models. The error bars are roughly the size of the symbols or smaller. Probability density functions (PDFs) of (e) stellar mass, (f) SFR, (g) gas-phase oxygen abundance and (h) dust attenuation optical depth derived with the three libraries using the same colour code as above. (i, j, k) SFHs of the best-fitting models (black solid lines), likelihood-weighted average SFHs (black dashed lines) and associated confidence ranges (likelihood-weighted standard deviation, shades) from the three fits. The residuals show that all three fits are reasonably good, but the PDFs reveal large differences in the extracted parameters. The PDFs derived using the CLSC spectral library in (g) and (h) look unrealistically narrow and hit the edge of the prior (see text for details). The original P12 library yields narrower PDFs than the P12nEL library. This illustrates how accounting for the contamination of broadband fluxes by emission lines can help constrain the parameters better (see Section~\ref{sec:resphot}). The SFHs estimated using the P12nEL (j) and P12 (k) spectral libraries rise as a function of time, in contrast to those derived using the CLSC spectral library (i).}
\label{fig:fitp}
\end{center}
\end{figure*}

\begin{figure*}
\begin{center}
\includegraphics[width=0.75\textwidth]{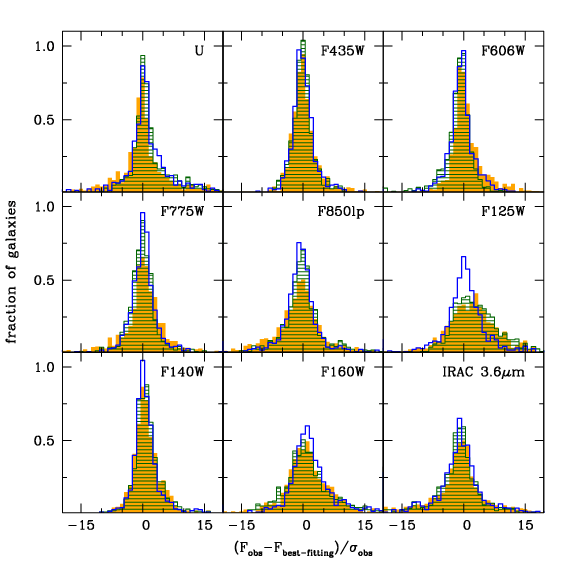}
\vspace{-0.25cm}
\caption{Distribution of fit residuals between the observed flux $F_\mathrm{obs}$ and best-fitting model flux $F_\mathrm{best-fitting}$, in units of the observational error $\sigma_\mathrm{obs}$, for all 1048 galaxies in the 3D-HST photometric sample. Each panel refers to a different photometric band (indicated in the top-right corner). The different histograms refer to the CLSC (yellow shaded histogram), the P12nEL (green hatched histogram) and the original P12 (blue solid histogram) model spectral libraries. We note that the residuals extend to large values because the quote photometric uncertainties on the observed fluxes are very small, of the order of 2 percent of the total flux (as discussed in Section~\ref{sec:sample}). The three different spectral libraries provide reasonable fits to the data, although the best results (i.e. narrowest histograms most centred on zero) are obtained with the P12 library.}
\label{fig:resid}
\end{center}
\end{figure*}

\begin{figure*}
\begin{center}
\includegraphics[width=0.47\textwidth]{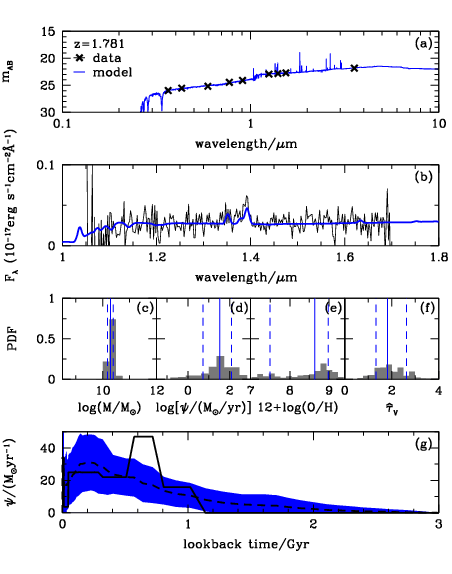}
\includegraphics[width=0.47\textwidth]{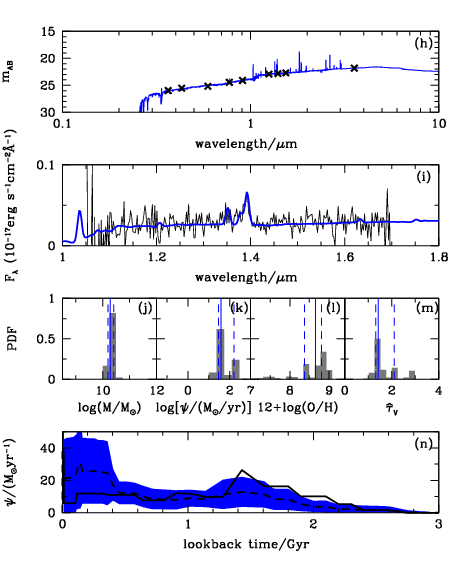}
\vspace{-0.25cm}
\caption{Best-fitting models and parameter PDFs for an example galaxy in the 3D-HST emission-line sample fitting the photometry alone (\textit{left-hand panel}) and fitting the photometry and the observed emission-line EWs simultaneously (\textit{right-hand panel}) with the P12 spectral library. (a, h) Observed SED (black crosses) and best-fitting models (blue solid lines). (b, i) Observed grism spectrum (black solid lines) and best-fitting model spectra at similar resolution (blue solid lines). PDFs of stellar mass (c, j), SFR (d, k), gas-phase oxygen abundance (e, l) and optical depth of the dust (f, m). 50th (solid line), 16th and 84th (dashed lines) percentiles of the PDFs are marked on top. (g, n) SFHs of the best-fitting models (black solid lines), likelihood-weighted average SFHs (black dashed lines) and associated confidence ranges (likelihood-weighted standard deviation, blue shades) in the two cases. The photometric fits look reasonable in both top panels, but the emission-line fluxes predicted when fitting the photometry alone do not match the observed ones. When including the emission-line EWs in the fit, the best-fitting model spectrum reproduces them satisfactorily and the constraints on the physical parameters are significantly tighter.}
\label{fig:fit}
\end{center}
\end{figure*}

We can quantify further the appropriateness of the three spectral modelling approaches investigated here to interpret photometric and spectroscopic observations of distant galaxies by comparing the constraints derived in each case on physical parameters, such as stellar mass, SFR and dust attenuation optical depth. To do so, we use the bayesian approach described in \cite{pacifici2012} (see also \citealt{kauffmann2003a,gallazzi2005,dacunha2008}). In brief, we compare the observational constraints (photometry$+$spectroscopy or photometry alone) on each 3D-HST galaxy to the same observable quantities predicted for each model in a given spectral library, such that $z_{\mathrm{model}}=z_{\mathrm{grism}} \pm 0.02$ (in practice, this implies that each observed galaxy is compared to about 65,000 models). We then use the likelihood of each model to build probability density functions (PDFs) of selected physical parameters for that galaxy: stellar mass ($M_{\ast}$); SFR ($\psi$); gas-phase oxygen abundance [$\aboh$]; and attenuation optical depth of the dust ($\tauv$).

For galaxies in the 3D-HST photometric sample, we fit the fluxes in all nine broad bands from 0.35 to 3.6$\mu$m.\footnote{Since the formal photometric errors can be extremely small (Section~\ref{sec:phot}), we adopt a minimum photometric error of 5 per cent to enlarge the number of models contributing to each fit.} In Fig.~\ref{fig:fitp}, we show an example of spectral fit of a galaxy at redshift $z=1.261$ using the three model spectral libraries described in Section~\ref{sec:model}. In Fig.~\ref{fig:fitp}a, we plot the observed magnitudes (black crosses) and the best-fitting model spectra from the CLSC (orange), P12nEL (green) and P12 (blue) spectral libraries. The residuals in Fig.~\ref{fig:fitp}bcd show that all three fits are reasonably good, but the PDFs plotted in Fig.~\ref{fig:fitp}efgh reveal large differences in the extracted physical parameters. Since all fits are performed in the same way, these discrepancies arise purely from differences in the model spectral libraries. Not only the median values, but also the widths of the PDFs are different in the three fits. In particular, the fit obtained using the CLSC spectral library shows narrow histograms of the gas-phase oxygen abundance (g) and the dust attenuation optical depth (h). If a narrow PDF can in some occasions reflect a good fit, this must be interpreted in the context of the number of fitted data points, the errors in the data, the distribution of the priors, the number of models in the spectral library and the number of models which actually contribute to the fit. In Fig.~\ref{fig:fitp}, differences in the derived PDFs of $M_{\ast}$ (e), $\psi$ (f), $\aboh$ (g) and $\tauv$ (h) using different libraries can arise only from differences in the prior distributions and the number of models contributing to each fit. In particular, the narrowness of the PDFs obtained using the CLSC library arises from the fact that only few models in this library are as blue as the observed galaxy (with both ACS and WFC3 colours close to zero in Fig.~\ref{fig:col3lib}), driving the fit to the most actively star-forming, most metal-poor and least attenuated models. The observed colours are better sampled by the other libraries, the original P12 library yielding narrower PDFs than the P12nEL library. This illustrates how accounting for the contamination of broadband fluxes by emission lines can help constrain these parameters better, as we discuss in Section~\ref{sec:resphot}. This point was emphasized by \cite{pacifici2012}, whose models were specifically designed to allow one to maximize the constraining power of photometric observations in the absence of direct spectroscopic information on emission lines. In Fig.~\ref{fig:fitp}ijk, we show the SFHs of the best-fitting models (black solid lines), the likelihood-weighted average SFHs (black dashed lines) and associated confidence ranges (likelihood-weighted standard deviation, shades) derived with the three spectral libraries. We note that the SFHs derived using the P12nEL (j) and P12 (k) spectral libraries rise as a function of time, in contrast to those derived using the CLSC spectral library.

To quantify in a more global way the ability of the different spectral libraries to account for the photometric properties of 3D-HST galaxies, we plot in Fig.~\ref{fig:resid} the distribution of the residuals between best-fitting model and observed fluxes, in each photometric band, for all 1048 galaxies in the sample. In each panel, the different histograms refer to the CLSC (yellow shaded histogram), the P12nEL (green hatched histogram) and the original P12 (blue solid histogram) model spectral libraries. Fig.~\ref{fig:resid} shows that, overall, the three different spectral libraries provide reasonable fits to the data, although as expected, the best results (i.e. narrowest histograms most centred on zero) are obtained with the P12 library. The CLSC spectral library shows slight systematic offsets in the $U$ band (where the best-fitting flux overestimates the observed one) and in the F606W and F775W bands (where the best-fitting flux underestimates the observed one). These discrepancies imply that the corresponding models do not reproduce the rest-frame ultraviolet-optical slopes of observed galaxies as well as those in the P12 spectral library, which we attribute to the oversimplified dust attenuation prescription and lack of stochasticity in the star formation histories in the CLSC spectral library. The most extended residual distributions in Fig.~\ref{fig:resid} pertain to the F125W band, in which the best-fitting fluxes in both the CLSC and P12nEL spectral libraries can severely underestimate the observed fluxes. This is because both models fail to reproduce the contamination of the F125W band flux by the {\ha} emission line. This happens at $z\lesssim1.2$, thus for more than half the galaxies in the photometric sample. Remarkably, when emission lines are included in the model spectra, as is the case for the P12 spectral library, the fit residuals improve significantly. We further discuss and quantify the contamination of observed broadband fluxes by emission lines in Section~\ref{sec:resspec}.

In the case of the 3D-HST emission-line sample, we can add to the constraints on the nine broad-band fluxes those on up to four emission-line EWs in the rest-frame optical wavelength range. Fig.~\ref{fig:fit} shows an example of fit (performed with the P12 spectral library) of such a galaxy at $z=1.781$ with available measurements of the {\hb} and {\oiii} emission lines. The left-hand panels show the results obtained when fitting the photometry alone and the right-hand panels those obtained when including also the constraints on the {\hb} and {\oiii} emission lines. In Fig.~\ref{fig:fit}ah, we show the observed SED (black crosses) together with the best-fitting models (blue solid line) and in Fig.~\ref{fig:fit}bi, the observed grism spectrum (black solid line) together with the best-fitting model spectra at similar resolution (blue solid line). The photometric fits performed with the P12 spectral library look reasonable in both top panels, but the emission-line fluxes predicted when fitting the photometry alone do not match the observed ones (b). When including the emission-line EWs in the fit, the best-fitting model spectrum reproduces them satisfactorily (i). In Fig.~\ref{fig:fit}, we show the constraints derived in both cases on the same physical parameters as in Fig.~\ref{fig:fitp}, i.e. stellar mass (c, j), SFR (d, k), gas-phase oxygen abundance (e, l) and attenuation optical depth of the dust (f, m). Except for the stellar mass (which is usually well constrained by multi-band photometry alone), the constraints obtained when including information on the {\hb} and {\oiii} emission lines (k, l, m) are significantly tighter than those derived from photometry alone (d, e, f). In particular for the SFR, the uncertainty decreases from 0.7 to 0.35 dex. This is expected, since the nebular emission lines arise from the obscured {\hii} regions ionized by young massive stars; thus measuring these lines allows one not only to probe current, massive star formation, but also the metallicity and dust content in those regions \citep{charlot2001,pacifici2012}. In Fig.~\ref{fig:fit}gn, we show the SFHs of the best-fitting models (black solid lines), the likelihood-weighted average SFHs (black dashed lines) and associated confidence ranges (likelihood-weighted standard deviation, blue shades) in the two cases. Including the emission-line equivalent widths in the fit improves the constraints on the \textit{current} (last 10 Myr) values of physical parameters, but generally does not affect the constraints on the full SFH.

In the next Section, we apply the above fitting procedure to all galaxies in the photometric and emission-line 3D-HST samples and inter-compare the results obtained with the different spectral libraries in terms of the accuracy and uncertainty of statistical constraints on physical parameters (stellar mass, SFR and attenuation optical depth of the dust) in a more global way.

\section{Statistical constraints on galaxy physical parameters}
\label{sec:results}

The accuracy and uncertainty of the constraints on galaxy physical parameters derived from statistical fits of observations depend both on the type of observations considered (photometry, spectroscopy) and on the model spectral library used to interpret these. In this Section, we first explore the effects of using the three different model spectral libraries of Section~\ref{sec:model}, which describe with different levels of sophistication the stellar and interstellar components of galaxies, to interpret the same set of observations (photometric 3D-HST sample). Then, we quantify the improvement introduced by the ability with the P12 spectral library to fit a combination of photometric and spectroscopic data (available for the 3D-HST emission-line sample) compared to fitting broad-band photometry alone. 

\subsection{Fits to the photometry: classical vs realistic models}
\label{sec:resphot}

\begin{figure*}
\begin{center}
\vspace{-0.9cm}
\includegraphics[width=0.95\textwidth]{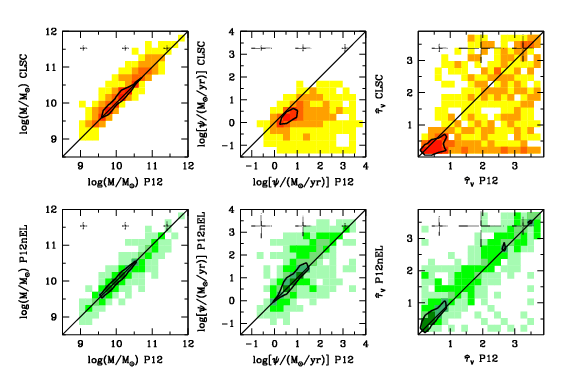}
\caption{Comparison between constraints of physical parameters derived for all 1048 galaxies in the 3D-HST photometric sample using the CLSC (top panels, in orange) and P12nEL (bottom panels, in green) model spectral libraries to those obtained using the more sophisticated P12 library. In each box, we show the comparison between the medians of the PDFs. The shades, from dark to light, mark 75, 50, 30, 10 and 1 per cent of the maximum density. The black contours mark 50 per cent of the maximum density. In the top part of each panel, we show also the average uncertainty in three bins. Each column represents a different physical parameter: stellar mass, SFR and optical depth of the dust, from left to right. The use of simple exponentially declining SFHs (CLSC spectral library) can cause strong biases on both the stellar mass and the SFR. The uncertainties on physical parameters estimated using this library may appear to be artificially narrow, hiding the fact that the models span a too narrow prior range. Not including the emission lines in the broad-band fluxes (P12nEL spectral library) does not strongly affect the estimates of stellar mass, but can induce a slight overestimation of the SFR.}
\label{fig:paramp}
\end{center}
\end{figure*}

\begin{table*}
\begin{center}
\caption{16, 50 and 84 percentiles of the distributions of the differences between best estimates and uncertainties of stellar mass, SFR and optical depth of the dust when comparing constraints with different libraries as shown in Fig.~\ref{fig:paramp}: CLSC vs P12 and P12nEL vs P12.}
\label{tab:diff}
\begin{tabular}{l c c c c c c | c c c c c c}
\hline
\hline
& \multicolumn{6}{c}{CLSC $-$ P12} & \multicolumn{6}{c}{P12nEL $-$ P12} \\
& \multicolumn{3}{c}{bias} & \multicolumn{3}{c}{$\Delta$ uncertainty} & \multicolumn{3}{c}{bias} & \multicolumn{3}{c}{$\Delta$ uncertainty} \\
& 16th & 50th & 84th &  16th & 50th & 84th & 16th & 50th & 84th & 16th & 50th & 84th  \\
\hline
$\log(M_{\ast}/M_{\sun})$ 				&{0.27}	& {0.08}	&{ -0.03}	&{0.02}	&{-0.02}	& {-0.07}	& 0.06	& 0.00	& -0.13	& 0.08	& 0.01	& -0.02 \\
$\log[\psi/(M_{\sun} \mathrm{yr}^{-1})]$ 	&{ -0.14}	& {-0.63}	& {-2.23}	&{0.03}	&{ -0.16}	& {-0.61}	& {0.62}	&{0.12}	& -0.15	&{0.35}	& {0.06}	& {-0.09} \\
$\hat{\tau}_V$ 						&{ 0.28}	& {-0.18}	& {-1.62}	&{0.26}	& {-0.03}	& {-0.41}	& {0.42}	& 0.07	& {-0.12}	& {0.17}	& 0.01	& {-0.16} \\
\hline
\hline
\end{tabular}
\end{center}
\end{table*}

In Fig.~\ref{fig:paramp}, we compare the constraints on stellar mass, SFR and optical depth of the dust derived for all 1048 galaxies in the 3D-HST photometric sample using the CLSC (top panels, in orange) and P12nEL (bottom panels, in green) model spectral libraries to those obtained using the more sophisticated P12 library. In all cases, the constraints are derived from fits of the 9-band photometry at observer-frame wavelengths between 0.35 and 3.6 $\mu$m, as described in Section~\ref{sec:fit}.

\paragraph*{Stellar mass.} The left-hand panels of Fig.~\ref{fig:paramp} show the differences in the constraints derived on galaxy stellar mass. For each set of spectral libraries, each panel shows the comparison between the median likelihood estimates of stellar mass (colour-coded according to the number of galaxies falling into each bin of the diagram). The results show that stellar-mass estimates derived using the CLSC spectral library are systematically $\sim0.08$ dex greater than those derived using the P12 library. This is because in the CLSC spectral library, simple exponentially declining SFHs tend to produce larger mass-to-light ratios and smaller specific SFRs than SFHs including late bursts of star formation. Such a trend can cause a bias in the mass estimate towards large values (although this depends on the allowed age range of the models; see below). Fig.~\ref{fig:paramp} also does not reveal any systematic bias in the stellar masses obtained using the P12nEL rather than the original P12 spectral library, only a large scatter. This suggests that stellar-mass estimates are not strongly affected by the contamination of broad-band fluxes by emission lines, at least for masses $M_{\ast}>10^9~\msun$.

\paragraph*{SFR.} The middle panels of Fig.~\ref{fig:paramp} show the analog to the left-hand panels for the SFR. In this case, the constraints derived using the CLSC spectral library produce SFR estimates significantly smaller (by $\sim 0.63$ dex) than those derived from the more sophisticated P12 library, again because the SFH prior favours low specific-SFR values. When comparing the constraints derived from the P12nEL and P12 spectral libraries, we see that neglecting nebular emission causes a slight bias towards larger SFRs ($\sim 0.12$ dex). This is likely because the total (stellar$+$nebular) observed emission is interpreted as a larger amount of young stars when emission lines are not included in the broad band fluxes predicted by the models. The uncertainties are also slightly larger when the emission lines are not included in the models.

\paragraph*{Dust optical depth.} The right-hand panels of Fig.~\ref{fig:paramp} pertain to the attenuation optical depth of the dust. In the CLSC spectral library, dust attenuation is computed using a two-component model with a fixed slope of the attenuation curve ($n=-0.7$) both in stellar birth clouds and in the diffuse ISM (as recalled in Section~\ref{sec:basic}, the galaxy-wide attenuation curve then depends on the SFH). The P12nEL and P12 spectral libraries rely on the same two-component model, but with a steeper slope of the attenuation curve in stellar birth clouds ($n=-1.3$) and a random slope of the attenuation curve in the diffuse ISM ($n$ between $-0.4$ and $-1.1$). On the one hand, this difference would cause a bias in the estimate of the optical depth of the dust, giving a larger value when applying a shallow attenuation curve. On the other hand, the large mass-to-light ratios and small SFRs favoured by the CLSC spectral library tend to produce redder intrinsic SEDs and favour low dust attenuation. These two effects conspire to produce a large scatter ($\pm$ 1 dex), and an offset ($\sim -$0.18 dex) between the results derived using the CLSC and P12 spectral libraries in the upper right panel of Fig.~\ref{fig:paramp}. The lower-right panel further shows that neglecting nebular emission in the P12 spectral library introduces a slight bias upward ($\sim$ 0.07) in the derived attenuation optical depth with a large scatter, showing the strong degeneracy of this parameter with age, metallicity and SFR. We note that, for some dusty galaxies ($\tauv>2.5$), the measurements of $\tauv$ derived with the P12nEL and P12 spectral libraries are in good agreement. This is because nebular emission lines are strongly attenuated at high dust optical depths and therefore have a negligible influence on the broad-band fluxes. As a consequence, in the lower-right panel of Fig.~\ref{fig:paramp}, the density along the identity relation is high when $\tauv$ is large.

\paragraph*{} We conclude from the above analysis that the widely used, CLSC spectral library is not appropriate to extract reliable stellar masses and SFRs from photometric observations of distant galaxies such as those in the 3D-HST survey, for two main reasons: firstly, because the limited parameter range probed by this library cannot account for all observations (as discussed in Sections~\ref{sec:mockobs} and \ref{sec:fit}); and secondly, because this discrepancy implies important systematic biases in derived galaxy physical parameters, even if the resulting PDFs can appear artificially narrow (see Section~\ref{sec:fit}). We note that the results presented here for the CLSC spectral library are consistent with the claim by \citet[see also \citealt{pforr2012}]{maraston2010} that models with exponentially declining SFHs overestimate the stellar mass and underestimate the SFR if galaxies are assumed to be at least $\sim1\,$Gyr old (see their section~3.2). The constraints on stellar mass, SFR and optical depth of the dust derived from 9-band photometry using the P12nEL and P12 spectral libraries are roughly similar, except for a slight bias in SFR toward large values when nebular emission is neglected. For reference, we list in Table~\ref{tab:diff} the 16th, 50th and 84th percentiles of the distributions of differences between the median-likelihood estimates of stellar mass, SFR and optical depth of the dust (and associated uncertainties) derived using the CLSC and P12nEL spectral libraries with respect to those derived using the more sophisticated P12 library.

\subsection{Including emission-line constraints in the spectral fits}
\label{sec:resspec}

\begin{figure*}
\begin{center}
\vspace{-0.9cm}
\includegraphics[width=0.95\textwidth]{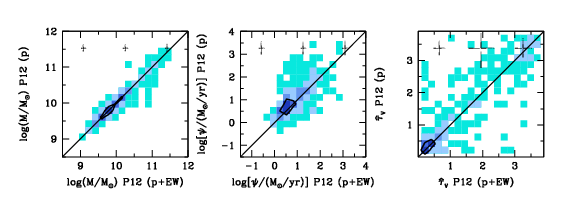}
\vspace{-0.25cm}
\caption{Comparison between the constraints derived on stellar mass, SFR and optical depth of the dust for all 364 galaxies in the 3D-HST emission-line sample when fitting the photometry alone and when fitting both the photometry and emission-line EWs, using the P12 model spectral library. The format is the same as in Fig.~\ref{fig:paramp}. Including the emission-line EWs in the fit does not affect strongly the best estimates, but improves considerably the uncertainties on all physical parameters.}
\label{fig:params}
\end{center}
\end{figure*}

\begin{table*}
\begin{center}
\caption{16, 50 and 84 percentiles of the distributions of the differences between best estimates and uncertainties of the stellar mass, SFR and attenuation optical depth of the dust as shown in Fig.~\ref{fig:params}, comparing simultaneous fits to the photometry and emission-line EWs vs. fits to the photometry alone (using the P12 model spectral library in both cases).}
\label{tab:diffew}
\begin{tabular}{l c c c c c c}
\hline
\hline
&  \multicolumn{6}{c}{P12 (phot) $-$ P12 (phot $+$ EW)}\\
&  \multicolumn{3}{c}{bias} & \multicolumn{3}{c}{$\Delta$ uncertainty}\\
&16th & 50th & 84th & 16th & 50th & 84th  \\
\hline
$\log(M_{\ast}/M_{\sun})$ 				& 0.09	& 0.01	& -0.07	& 0.07	& 0.03	& -0.01\\
$\log[\psi/(M_{\sun} \mathrm{yr}^{-1})]$ 	& 0.50	& {0.13}	& {-0.21}	& {0.48}	& {0.14}	& 0.02\\
$\hat{\tau}_V$ 						& {0.45}	& {0.03}	& {-0.23}	& {0.35}	& {0.06}	& {-0.09}\\
\hline
\hline
\end{tabular}
\end{center}
\end{table*}

\begin{figure*}
\begin{center}
\includegraphics[width=0.475\textwidth]{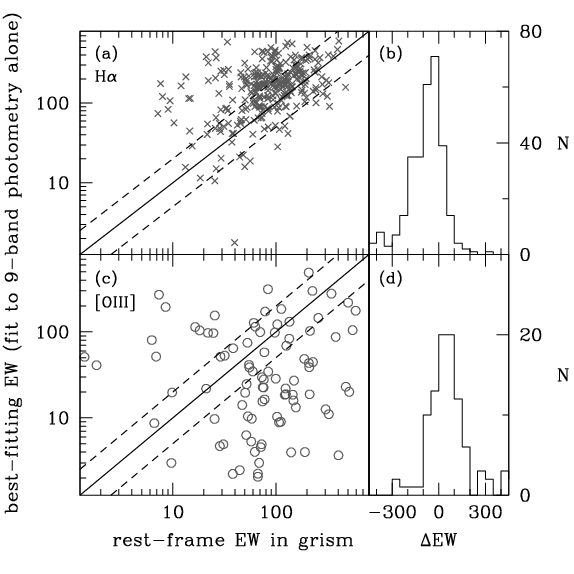}
\vspace{-0.5cm}
\caption{(a): comparison between the rest-frame emission EW of {\ha}  measured from grism spectroscopy and the EW predicted from fits of the 9-band photometry using the P12 model spectral library, for the 314 galaxies with \ha\ measurements in the 3D-HST emission-line sample. For reference, the black solid line indicates the identity relation and the black dashed lines mark deviations by a factor of 2 ($\pm0.3\,$dex) between the measured and predicted EWs. (b): distribution of the difference between predicted and measured \ha\ rest-frame EW. (c) and (d):  same as (a) and (b), but for the EW of \oiii\ for the 99 galaxies with \oiii\ measurements in the 3D-HST emission-line sample.}
\label{fig:ewmvsbfit}
\end{center}
\end{figure*}

\begin{figure*}
\begin{center}
\includegraphics[width=0.475\textwidth]{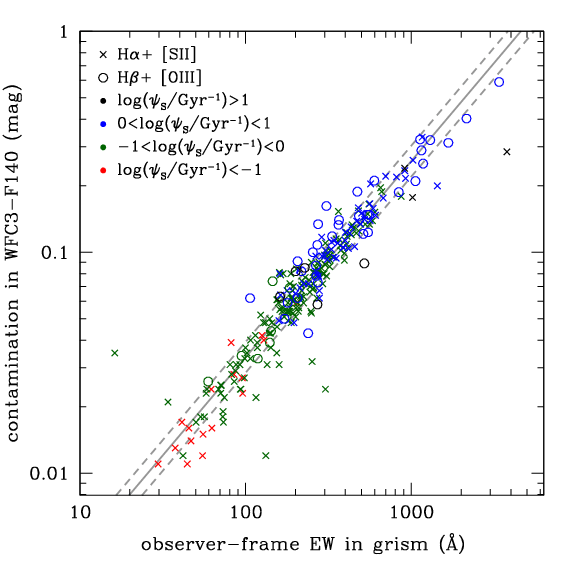}
\includegraphics[width=0.475\textwidth]{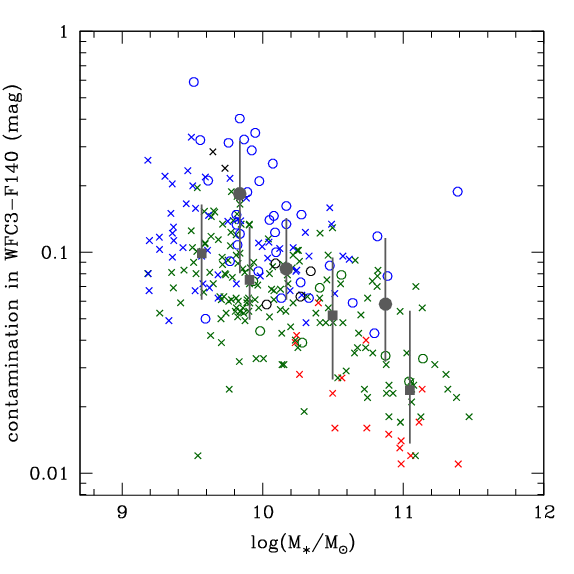}
\vspace{-0.5cm}
\caption{Contribution by nebular emission to the WFC3-F140W magnitude of the best-fitting (P12) model for the galaxies in the 3D-HST emission-line sample as a function of the observer-frame line EWs directly measured in the grism spectra (left-hand panel) and as a function of stellar mass (right-hand panel). Crosses represent galaxies with measured {\ha} and {\sii} (235 galaxies at $0.8<z<1.4$), while circles represent galaxies with measured {\hb} and {\oiii} (47 galaxies at $1.5<z<2.2$). The points are colour-coded according to star formation activity from low (red) to high (black) specific SFR. In the left-hand panel, the grey lines show the same relation considering the best-fitting model EWs instead of the observed ones. In the right-hand panel, grey symbols represent median and 16-to-84 percentiles of the contamination in bins ($\approx 0.5$ dex wide) of stellar mass for {\ha} and {\sii} (squares) and for {\hb} and {\oiii} (filled circles). The contamination of the emission lines in the broad-band WFC3-F140W filter increases from $\approx 0.01$ to $\approx 0.5\,$mag as the stellar mass decreases.}
\label{fig:ewband}
\end{center}
\end{figure*}

The 3D-HST emission-line sample includes only those galaxies from the photometric sample with well-detected (S/N~$>5$) emission lines, which allows us to test the effect of adding emission-line information when deriving constraints on physical parameters. In Fig.~\ref{fig:params}, we compare (in a similar way to Fig.~\ref{fig:paramp}) the constraints derived on stellar mass, SFR and optical depth of the dust for all 364 galaxies in this sample when fitting the photometry alone and when fitting both the photometry and emission-line EWs, using the P12 model spectral library. For the stellar mass (left-hand panel), there is no significant difference between the constraints derived in both ways and the uncertainties remain similar. The SFR (middle panel) appears to be slightly overestimated when the constraints on the emission-line EWs are not included in the fit (by $\sim0.13$~dex). This is because emission-line EWs help break the degeneracy between star formation activity (i.e. specific SFR) and attenuation by dust (right-hand panel in Fig.~\ref{fig:params}) at fixed SED shape. The fact that the SFR is biased high when fitting only the photometry can be considered as a consequence of the selection criteria of the emission line sample, because, for an emission line to be well detected in a low-resolution 3D-HST spectrum, attenuation by dust must be low. In the absence of emission-line information, $\tauv$ is less well constrained and the corresponding PDF broadens to larger values (implying larger SFR at fixed observed colours). The introduction of line-EW constraints breaks the SFR-dust degeneracy in these lightly obscured galaxies therefore leads on average to slightly lower SFR estimates. Adding constraints on emission-line EWs reduces the uncertainties in both SFR and $\tauv$ by $\sim0.14$\,dex and 0.06, respectively (Table~\ref{tab:diffew}).

It is important to note that, for galaxies in the 3D-HST emission-line sample, the emission-line strengths predicted by the P12 model providing the best fit to 9-band photometry alone are in fair agreement with direct EW measurements from grism spectroscopy. This is illustrated by Fig.~\ref{fig:ewmvsbfit}, in which we compare the rest-frame EWs of {\ha} (Fig.~\ref{fig:ewmvsbfit}a) and {\oiii} (Fig.~\ref{fig:ewmvsbfit}c) emission lines as measured from the grism spectra with the predictions obtained when fitting the photometry alone. The distributions of the differences between predicted and measured EWs are plotted in Figs~\ref{fig:ewmvsbfit}b and \ref{fig:ewmvsbfit}d for {\ha} and {\oiii}, respectively. Fig.~\ref{fig:ewmvsbfit} shows that, for 52 per cent (20 per cent) of the galaxies with spectroscopic {\ha} ({\oiii}) emission-line measurements, the EWs inferred from fits of 9-band photometry agree to within a factor of 2 with those measurements. Considering only galaxies with EW({\ha}) larger than 100\,{\AA} makes this fraction rise to 73 per cent. This is not surprising, as the P12 model spectral library accounts for contamination of broad-band fluxes by nebular emission, which increases and hence is more easily identifiable in the most actively star-forming galaxies. We note that, despite this good agreement, 9-band photometric fits tend to systematically overestimate EW(\ha) in Fig.~\ref{fig:ewmvsbfit}a relative to direct measurements. This is again a consequence of the SFR-dust degeneracy described in the previous paragraph. There is no systematic bias in the predicted EW(\oiii) but a larger scatter, because the strength of this line is strongly affected by the metallicity of the gas and is not easily constrained by purely photometric fits.

To further quantify how emission lines contaminate observed broad-band fluxes, we record, for each galaxy in the 3D-HST emission-line sample, the contribution by nebular emission to the WFC3-F140W magnitude of the best-fitting P12 model (as derived when including the constraints from both 9-band photometry and EW measurements). This is shown in Fig.~\ref{fig:ewband} (left-hand panel) against the line EWs directly measured in the grism spectra. For galaxies at redshifts $0.8 < z < 1.4$, both {\ha} and {\sii} fall in the WFC3-F140W filter. The combined observer-frame EW of these lines is plotted in Fig.~\ref{fig:ewband} (crosses). In the same way, we plot the combined observer-frame EW of {\hb} and {\oiii} for galaxies in the range $1.5 < z < 2.2$ (empty circles). Each galaxy is colour-coded according to star formation activity, from low (red) to high (black) specific SFR. As expected, the contamination of broad-band fluxes by emission lines increases with the level of star formation activity. The bulk of the sample shows observed emission-line EWs between 200 and 700~\AA, which corresponds roughly to a contamination of $0.1\,$magnitudes in the F140W broad-band magnitude. We also show for reference the relation obtained when considering the EWs of the best-fitting model instead of the observed ones (grey lines). In the right-hand panel of Fig.~\ref{fig:ewband}, we plot the emission-line contamination of the WFC3-F140W broad-band magnitude as a function of stellar mass. For galaxies at $0.8 < z < 1.4$, where {\ha} and {\sii} are both sampled in the band, the contamination decreases from $\approx 0.1$ to $\approx 0.02\,$magnitudes as the stellar mass increases from $\approx 10^{9.5}$ to $\approx 10^{11}$  M$_{\odot}$.  At higher redshift, where {\oiii} and {\hb} are sampled in the band, the contamination is slightly larger because the SFR is on average larger at higher than at lower redshifts \citep{noeske2007}.

\section{Implications for the scatter in the star-formation main sequence}
\label{sec:ms}

\begin{figure*}
\begin{center}
\includegraphics[width=0.8\textwidth]{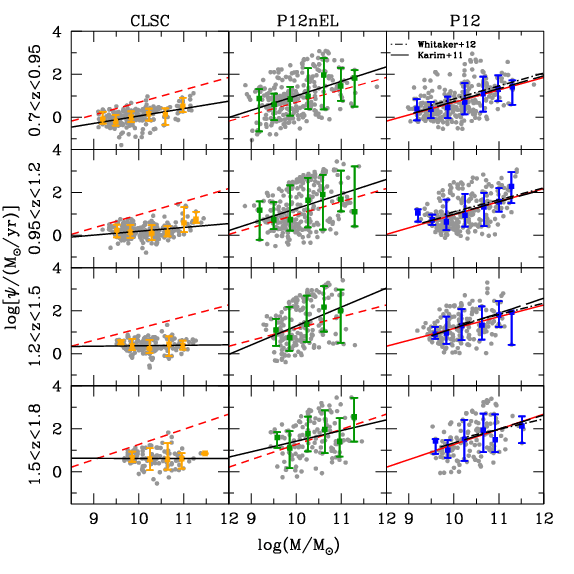}
\vspace{-0.5cm}
\caption{Constraints on stellar mass and SFR (star-formation main sequence) derived in Section~\ref{sec:resphot} from fits of the 9-band 3D-HST photometry using the CLSC (left-hand panels), P12nEL (middle panels) and P12 (right-hand panels) model spectral libraries, in different redshift ranges (rows). In each panel, we show the median and 16-to-84 percentile range of the SFR in bins of stellar mass (coloured squares) and linear least-squares fit to all points (black solid line; red solid line in the right-hand panels). For each redshift bin, we report in the 2 left-most panels the fit obtained in the right-hand panel using the P12 spectral library (red dashed line). The use of the CLSC spectral library, which relies on oversimplified prescriptions of SFHs and dust attenuation, cause the main sequence to lie, at all redshifts, significantly below the relation obtained using the more realistic P12 spectral library. The relative zero-point of the main sequence derived using the P12nEL spectral library is shifted towards higher SFRs compared to that obtained using the original P12 spectral library. The scatter about the derived star formation main sequence is reduced including the emission lines as contaminants of the broad-band fluxes. In the right-hand panel, we also report for comparisons the fits to the main-sequence by \protect\citet[black long-dashed line]{karim2011} and \protect\citet[black dot-dashed line]{whitaker2012} in the mass ranges allowed by their samples.}
\label{fig:msall}
\end{center}
\end{figure*}

\begin{figure*}
\begin{center}
\includegraphics[width=0.8\textwidth]{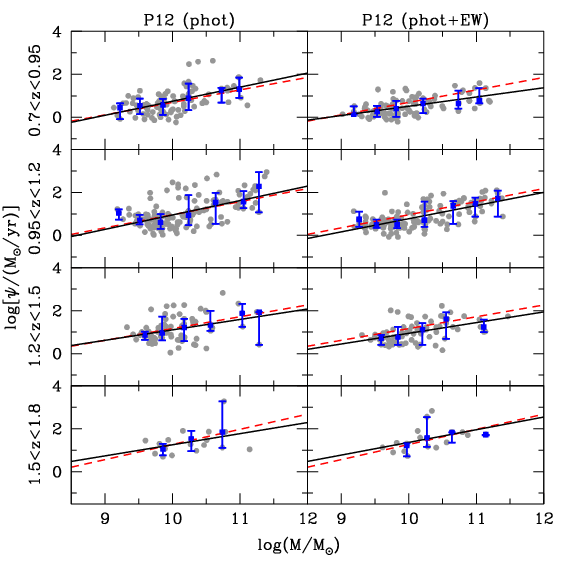}
\vspace{-0.5cm}
\caption{The main sequence derived from the 3D-HST emission-line sample when fitting the photometry alone (left-hand panels; analog of the right-hand panels of Fig.~\ref{fig:msall}) and when conbining the photometry with the emission-line EWs detected in the grism spectra (right-hand panels). The format is the same as in Fig.~\ref{fig:msall}. Estimates are extracted using the P12 spectral library. For this sample, the estimates of SFR are in general less uncertain than for the photometric sample given the strength of the emission lines. This contributes to reducing the scatter. Including the information from the emission-line EWs further reduces the scatter at all redshifts.}
\label{fig:ms}
\end{center}
\end{figure*}

We have shown in the previous sections that interpreting the same 3D-HST observations of distant galaxies using model spectral libraries of different levels of sophistication can lead to different constraints on galaxy stellar masses and SFRs. We also showed that accounting for nebular emission is important to interpret (even purely photometric) observations of distant star-forming galaxies. Here we explore how the biases introduced by the use of oversimplified model spectral libraries might affect the interpretation of the correlation between stellar mass and SFR observed for star-forming galaxies, commonly referred to as the star-formation main sequence. It is now established that this `main sequence' is in place both in the local Universe and at high redshift (up to at least $z\approx3$; \citealt{brinchmann2004,noeske2007,elbaz2007,daddi2007,karim2011,reddy2012,whitaker2012}), but the normalization, slope and scatter of the relation and the dependence of these on redshift, are still under debate. The normalization of the main sequence appears to increase with time (from redshift 0 to roughly 3) in such a way that, at fixed stellar mass, galaxies at higher redshift form stars at higher rates. The slope also appears to change with redshift, as shown by \cite{karim2011} and \cite{whitaker2012}. The intrinsic scatter should contain important information about the physical processes that drive the formation of galaxies \citep{dutton2010}, but it is always contaminated by observational uncertainties and thus is hard to quantify \citep{guo2013}.

In this Section, we assess how the use of different model spectral libraries and the availability or not of emission-line measurements can affect the determinations of the shape and scatter of the main sequence at $0.7<z<1.8.$\footnote{Because of the lack of statistics, we limit the analysis of the star-formation main sequence to $z=1.8$. We note also that the samples we use reach a magnitude of $H=23$ (WFC3-F160W) and thus are complete only to a stellar mass of $\approx10^{10}\msun$ at $z=1$ ($\approx10^{10.7}\msun$ at $z=2$). Such completeness limits do not allow us to draw reliable conclusions about the absolute slope or the zero point of the main sequence. It is thus beyond the scope of this paper to assess the evolution of the main sequence in redshift.} We use the estimates of stellar mass and SFR derived in Section~\ref{sec:results} to draw the star-formation main sequence in four redshift bins for both the 3D-HST photometric and the emission-line samples.\footnote{We include only galaxies with specific SFR larger than 0.01 Gyr$^{-1}$.} Fig.~\ref{fig:msall} shows the results obtained when adopting the constraints on stellar mass and SFR derived in Section~\ref{sec:resphot} from fits of the 9-band 3D-HST photometry using the CLSC (left-hand panels), P12nEL (middle panels) and P12 (right-hand panels) model spectral libraries. In each panel, on top of the single points pertaining to individual galaxies, we plot the median and 16-to-84 percentile range of the SFR in bins of stellar mass (coloured squares) and a linear least-squares fit to all points (black solid line in the 2 left-most panels; red solid line in the right-hand panel). For the purpose of comparison, for each redshift bin, we report in the 2 left-most panels the fit obtained in the right-hand panel using the P12 spectral library (as a red dashed line). Fig.~\ref{fig:msall} confirms our finding in Section~\ref{sec:resphot} above that the use of the CLSC spectral library, which relies on oversimplified prescriptions of SFHs and dust attenuation, can severely bias estimates of galaxy stellar masses and SFRs from photometric data. Such biases cause the main sequence to lie, at all redshifts, significantly below the relation obtained using the more realistic P12 spectral library. The primary cause for this can be traced to the prior distribution of the specific SFR, which in the CLSC spectral library is relatively narrow with a peak around $\log(\psi_{S}/\mathrm{Gyr}^{-1})\sim -0.75$.

The middle panels of Fig.~\ref{fig:msall} show the results obtained when including more sophisticated treatments of star formation and chemical enrichment histories and attenuation by dust, but without accounting for nebular emission (P12nEL spectral library). This produces a stronger correlation between stellar mass and SFR than obtained with the CLSC library, but with a very large scatter spanning almost the entire range of the prior distribution of the specific SFR ($-2<\log(\psi_{S}/\mathrm{Gyr}^{-1})<2$). This is caused by the strong degeneracy between SFR and dust attenuation, which affects the uncertainties in SFR estimates at all redshifts (Section~\ref{sec:resphot}). We also find that the relative zero-point of the main sequence derived using the P12nEL spectral library is shifted towards higher SFRs compared to that obtained using the original P12 spectral library. This is in agreement with the result of Fig.~\ref{fig:paramp} that the SFR is slightly overestimated when the models do not account for the contamination of broadband fluxes by emission lines.

Appealing to the original P12 library to include the contamination of broadband fluxes by nebular emission reduces the scatter about the derived star formation main sequence ($\approx0.7$~dex), especially at those redshifts where the {\ha} line falls in one of the observer-frame near-infrared bands ($0.7<z<1.5$); at $z>1.5$, the scatter remains very large ($\approx1.2$~dex) because the parameters are not as tightly constrained.

The star-formation main sequence has been studied in the past using different approaches tailored to specific redshift ranges and datasets. For comparison, we plot in the right-hand panels of Fig.~\ref{fig:msall} the main-sequence fits derived by \citet[black long-dashed line]{karim2011} and \citet [black dot-dashed line]{whitaker2012} over a redshift range similar to that sampled by the 3D-HST survey. In both studies, stellar masses are derived using a classical approach (exponentially declining $\tau$-models, possibly allowing for very young ages to widen the specific-SFR and mass-to-light-ratio priors), while SFRs are derived in different ways: \cite{karim2011} estimate galaxy SFRs from stacked 1.4 GHz data using the prescription by \cite{bell2003}; and \cite{whitaker2012} from ultraviolet and infrared data using the prescription by \cite{kennicutt1998}. The main sequences obtained in these two studies are both in good agreement with that derived from 3D-HST data using the P12 library in Fig.~\ref{fig:msall}. An advantage of the approach presented here is the ability to derive, for each galaxy, simultaneous constraints on the stellar mass and SFR from a library of star formation and chemical enrichment histories, which allows one to also constrain other physical parameters (age, dust attenuation, metallicity).

We can now turn to the 3D-HST emission-line sample and assess whether the addition of spectroscopic emission-line measurements can improve determinations of the star formation main sequence with the P12 model spectral library. Fig.~\ref{fig:ms} shows the main sequences derived for this sample in the same redshift bins as for the photometric sample in Fig.~\ref{fig:msall}, both from fits of the photometry alone (left-hand panels; this is the analog of the right-hand panels of Fig.~\ref{fig:msall}, but for the sub-sample of 3D-HST galaxies with emission-line measurements) and from combined fits of the photometry and emission-line EWs (right-hand panels). Even when fitting the photometry alone, the scatter in Fig.~\ref{fig:ms} is, to some extent, reduced ($\approx 0.5$~dex) compared to that in the right-hand panels of Fig.~\ref{fig:msall}. This is because the average uncertainty in the SFR derived from 9-band photometry  in the emission-line sample (which includes only galaxies with well-detected emission lines) is slightly smaller than that for the entire photometric sample and so is the scatter in the inferred relation between stellar mass and SFR.  When we add the constraints on the emission-line EWs, the uncertainty on the SFR is further reduced (Section~\ref{sec:resspec}) and the main sequence becomes tighter at all redshifts (with a scatter $\approx 0.4$ dex; right-hand panels in Fig.~\ref{fig:ms}). The slight bias towards lower SFRs at large stellar masses when including the emission-line EWs in the fits is associated to the selection of the sample as discussed in Section~\ref{sec:resspec}.

\section{Summary and conclusions}
\label{sec:summary}

Interpreting ultraviolet-to-infrared observations of distant galaxies in terms of constraints on physical parameters -- such as stellar mass, SFR and attenuation by dust -- requires spectral synthesis modelling. In this paper, we have investigated how increasing the level of sophistication of standard simplifying assumptions of such models can improve estimates of galaxy physical parameters. To achieve this, we have compiled a sample of 1048 galaxies at $0.7<z<2.8$ with accurate photometry at rest-frame ultraviolet to near-infrared wavelengths from the 3D-HST survey, 364 of which have strong enough optical-line emission to be measured with good S/N from grism spectroscopy. Such spectroscopic information is important to allow us to quantify, from a pure observational point of view, the contamination of broadband photometric fluxes by emission lines. We have compared the SEDs of the galaxies in this sample with those from different model spectral libraries (of different levels of sophistication) to derive bayesian estimates of stellar mass, star formation rate and optical depth of the dust. We have found the following.
\begin{enumerate}
\item Classical spectral libraries, in which galaxy SFHs are assumed to be exponentially declining functions of time ($\tau$-models), cannot reproduce the observed SEDs of all galaxies in the sample, because such libraries do not account for the colours of galaxies with old populations undergoing a new starburst. \cite{maraston2010} and \cite{pforr2012} introduced exponentially {\em rising} SFHs to overcome this limitation, but such parametrization cannot reproduce the potential rise and fall of galaxy SFHs. The more realistic SFHs implemented in the P12 library \citep{pacifici2012}, which rely on the versatile post-treatment of cosmological simulations, allow us to adequately reproduce the observed colours of all galaxies in the 3D-HST sample.
\item As a result, stellar masses derived using classical spectral libraries (generally associated with an oversimplified prescription for dust attenuation and the neglect of nebular emission) tend to be systematically overestimated (by a median $\sim 0.1$ dex), and SFRs systematically underestimated (by a median $\sim 0.6\,$dex), relative to the values derived adopting the more realistic P12 spectral library (which includes a sophisticated and versatile prescription for dust attenuation as well as a careful account for nebular emission). Moreover, the uncertainties derived using classical spectral libraries can be misleadingly small because of the inappropriate sampling of the physical parameter space (Section~\ref{sec:fit}).
\item Neglecting the contamination of broadband fluxes by emission lines, while using the same SFHs and dust prescription as in the P12 spectral library, would lead to systematic SFR overestimates (by a median $\sim 0.1\,$dex), leaving stellar-mass estimates nearly unaffected. We note that galaxies in the 3D-HST sample have masses $M \ga10^9\msun$. In lower-mass galaxies with high specific SFR, we expect SFR and stellar-mass estimates obtained when neglecting nebular emission to be more severely biased than found here \citep[e.g.][]{atek2011,stark2013}.
\item The simultaneous fit of photometric broad-band fluxes and emission-line EWs, as enabled by the \cite{pacifici2012} approach, helps break a fundamental degeneracy between SFR and dust attenuation, which reduces considerably the uncertainties in these derived parameters (by $\sim0.1$--0.5\,dex; the uncertainty in stellar-mass estimates dropping by only $\sim0.05\,$dex).
\end{enumerate}

The results obtained in this paper reveal the importance of choosing appropriate spectral models to interpret deep galaxy observations. In particular, the biases introduced by the use of classical spectral libraries to derive estimates of SFR, stellar mass and attenuation by dust in distant galaxies can significantly affect the interpretation of standard diagnostics diagrams of galaxy evolution. These include, for example, the galaxy stellar-mass function \citep[see also][]{marchesini2009,muzzin2013}; the main sequence of star-forming galaxies and its intrinsic scatter; and the relations between stellar mass, SFR and other galaxy physical parameters, such as stellar age and metallicity \citep[e.g.][]{mannucci2010}. In this context, the spectral library developed by \cite{pacifici2012} offers the possibility to interpret these and other fundamental diagnostics on the basis of more realistic, and at the same time more versatile models. This is all the more valuable in that the approach can be straightforwardly tailored to the analysis of any combination of photometric and spectroscopic observations of galaxies at any redshift.


\section*{acknowledgements}

We thank the Referee for the very useful report that helped improving the Paper. This work was funded in part by the Marie Curie Initial Training Network ELIXIR of the European Commission under contract PITN-GA-2008-214227 and in part by the KASI-Yonsei Joint Research Program for the Frontiers of Astronomy and Space Science funded by the Korea Astronomy and Space Science Institute. E.d.C. acknowledges funding through the ERC grant `Cosmic Dawn'. S.C. acknowledges support from the European Research Council via an Advanced Grant under grant agreement no. 321323-NEOGAL. S.K.Y was supported by the National Research Foundation of Korea (Doyak 2014003730).


\def\aj{AJ}
\def\araa{ARA\&A}
\def\apj{ApJ}
\def\apjl{ApJ}
\def\apjs{ApJS}
\def\apss{Ap\&SS}
\def\aap{A\&A}
\def\aapr{A\&A~Rev.}
\def\aaps{A\&AS}
\def\mnras{MNRAS}
\def\pasp{PASP}
\def\pasj{PASJ}
\def\qjras{QJRAS}
\def\nat{Nature}

\def\aplett{Astrophys.~Lett.}
\def\aas{AAS}
\let\astap=\aap
\let\apjlett=\apjl
\let\apjsupp=\apjs
\let\applopt=\ao

\bibliographystyle{mn2e}
\bibliography{bib_pacifici}

\section*{Appendix A. Comparison of the constraints on physical parameters derived using the P12 approach and the code FAST}
\renewcommand\thefigure{A\arabic{figure}}
\setcounter{figure}{0}

\begin{figure*}
\begin{center}
\includegraphics[width=0.9\textwidth]{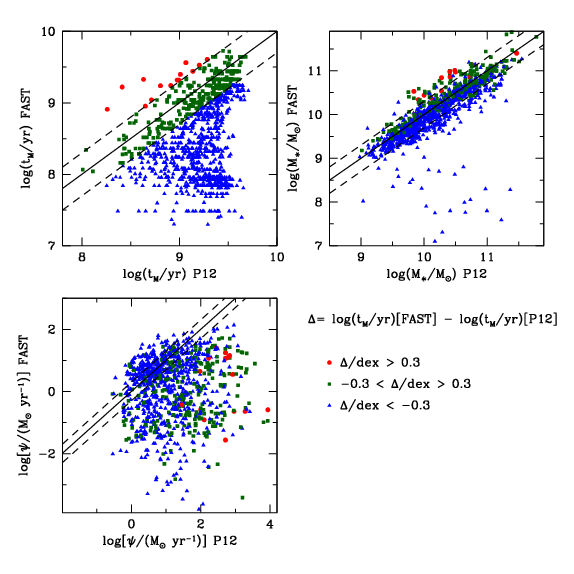}
\vspace{-0.5cm}
\caption{Estimates of mass-weighted age (a), stellar mass (b) and SFR (c) derived using the P12 approach, compared to those derived by the code FAST, in which the spectral library is built considering exponentially declining SFHs, fixed solar metallicity, \protect\cite{calzetti2000} dust prescription and no nebular emission. In each panel, symbols represent the galaxies in the 3D-HST photometric sample, the solid line marks the identity relation and the dashed lines deviations of a factor of 2 (0.3 dex) about the identity. Galaxies are colour-coded by the difference in age estimates between the two methods.}
\label{fig:app:mmsfr}
\end{center}
\end{figure*}

The main aim of this paper is to investigate how the level of sophistication of model spectral libraries adopted to fit galaxy observations can affect estimates of selected physical parameters. In the main part of the paper, to best illustrate the potential biases introduced by the choice of a given spectral library, we employ everywhere the same bayesian fitting technique to derive galaxy physical parameters. It is also worth comparing the constraints derived using the P12 approach (as explained in Sections~\ref{sec:model} and \ref{sec:fit}) and a different code, FAST \citep{kriek2009}, which has been used to constrain the properties of 3D-HST galaxies in many studies (among the most recent: \citealt{vanderwel2014}; \citealt{vandokkum2014}; \citealt{lundgren2014}; \citealt{patel2013}; \citealt{price2014}; and \citealt{fumagalli2013}). FAST is an IDL-based code, which fits stellar population synthesis templates to broadband photometry and/or spectra. The grid of models adopted in FAST to fit the sample of 3D-HST galaxies consists of exponentially declining SFHs ($e$-folding timescale in the range $0.01<\tau/\mathrm{Gyr}<10$) at fixed (solar) metallicity, spanning the age range from 0.04 to 12.6 Gyr, combined with \cite{bruzual2003} stellar population synthesis models for a \cite{chabrier2003} IMF, adopting the \cite{calzetti2000} dust prescription with $V$-band extinction in the range $0<A_V<4$.

In Fig.~\ref{fig:app:mmsfr}, we compare the estimates of mass-weighted age, stellar mass and SFR extracted using the P12 approach (as described in Section~\ref{sec:fit}) and FAST. In each panel, the symbols represent galaxies in the 3D-HST photometric sample, colour-coded according to the difference in age estimates between the two methods. The solid line marks the identity relation and the dashed lines deviations of a factor of 2 (0.3\,dex) about the identity. In Fig.~\ref{fig:app:mmsfr}a, we compare the estimates of mass-weighted age. For 26 per cent of the galaxies in the photometric sample, the estimates obtained using both approaches agree to within a factor of 2 (green squares). For 73 per cent of the galaxies, the age derived using FAST is more than a factor of two younger than that derived using the more sophisticated P12 approach (blue triangles). This is interesting because age estimates depend strongly on the chosen parameterization of the SFH: a library based on exponentially declining $\tau$-models must include very young (sometimes unrealistic) ages in order to reach high-enough specific SFRs and reproduce the colours of the bluest galaxies in the 3D-HST sample. Red circles mark the remaining 1 per cent of the galaxies (i.e., the few cases where the age derived by FAST is more than a factor of two older than that derived using the P12 approach). The difference in stellar-mass estimates (Fig.~\ref{fig:app:mmsfr}b) has a scatter of $\approx 0.4\,$dex, although differences for individual galaxies can reach up to 2--3 dex in some cases. We find that this scatter roughly correlates with the difference in age estimates: FAST stellar masses are larger when ages are older compared to those derived with the P12 approach (red circles); vice versa, the stellar masses derived by FAST are smaller, when the ages are younger (blue triangles). In Fig.~\ref{fig:app:mmsfr}c, we compare the SFR estimates. The estimates obtained using the two approaches agree to within $\pm 0.3$ dex for $\sim 45$ per cent of the galaxies in the 3D-HST sample, while  for most other galaxies, FAST-derived SFRs tend to be smaller than P12-derived ones. This bias in SFR estimates can be caused by the lack of stochasticity in the SFHs adopted by FAST for this analysis. We note that, as pointed out by \cite{wuyts2011} and \cite{price2014}, the inclusion of very short $e$-folding star formation timescales ($\tau<0.3$ Gyr) can also cause the SFR to be underestimated.

Overall, the comparison in Fig.~\ref{fig:app:mmsfr} is similar to that between the constraints derived using the P12 and CLSC spectral libraries in Section~\ref{sec:resphot} (Fig.~\ref{fig:paramp}). The main difference is that FAST-derived stellar masses can be underestimated relative to P12-derived ones in Fig.~\ref{fig:app:mmsfr}, while this was not the case for CLSC-derived stellar masses in Fig.~\ref{fig:paramp}. This is because FAST allows galaxy ages as young as 40\,Myr, while in the CLSC spectral library, the oldest stellar component of a galaxy is at least 1\,Gyr old.

\section*{Appendix B. Comparison between different versions of the Bruzual~\&~Charlot stellar population synthesis models}
\renewcommand\thefigure{B\arabic{figure}}
\setcounter{figure}{0}

\begin{figure}
\begin{center}
\includegraphics[width=0.47\textwidth]{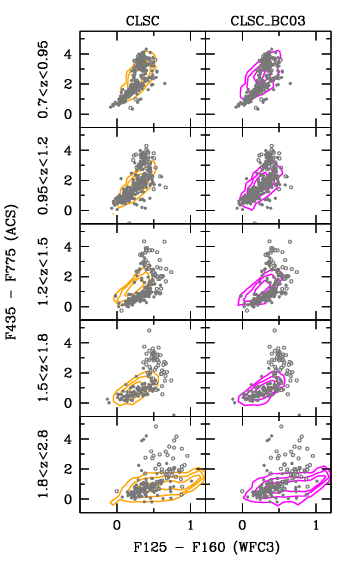}
\vspace{-0.3cm}
\caption{Optical-NIR colour-colour diagrams comparing the 3D-HST photometric sample with two spectral libraries. The format is the same as in Fig.~\ref{fig:col3lib}. The CLSC spectral library (left-hand side, orange) is identical to the one described in Section~\ref{sec:basic}. To build the CLSC\_BC03 spectral library (right-hand side, magenta) we adopt the same SFH and dust-attenuation prescriptions as for the CLSC spectral library, but we use the original version of the \protect\cite{bruzual2003} stellar models.}
\label{fig:app:bc03}
\end{center}
\end{figure}

\begin{figure}
\begin{center}
\includegraphics[width=0.47\textwidth]{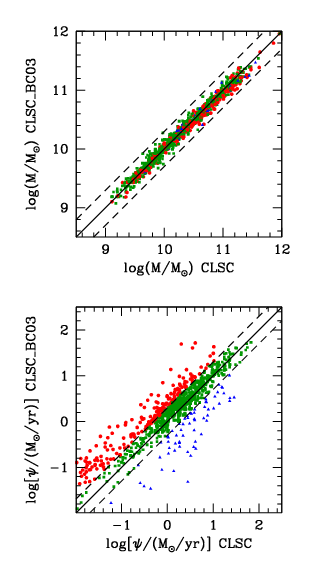}
\vspace{-0.3cm}
\caption{Comparison between the estimates of stellar mass (top panel) and SFR (bottom panel) derived with the CLSC and the CLSC\_BC03 spectral libraries. In each panel, symbols represent the galaxies in the 3D-HST photometric sample, the solid line marks the identity relation and the dashed lines deviations of a factor of 2 (0.3 dex) about the identity. Symbols are colour-coded according to the difference between the estimates of SFR: blue triangles (red circles) mark galaxies for which the estimates derived with the CLSC\_BC03 spectral library are smaller (larger) than those derived with the CLSC spectral library by more than 0.3 dex (6 and 22 percent of the galaxies respectively); green squares mark galaxies for which the estimates are in agreement within $\pm 0.3$ dex (72 percent of the galaxies).}
\label{fig:app:bc03param}
\end{center}
\end{figure}

Throughout this Paper, we adopt an updated version of the \cite{bruzual2003} stellar population synthesis models. This version incorporates a new library of observed stellar spectra \citep{sanchez2006} and new prescriptions for the evolution of stars less massive than 20 $M_{\odot}$ \citep{bertelli2008,bertelli2009} and for the thermally pulsating asymptotic giant branch (TP-AGB) evolution of low- and intermediate-mass stars \citep{marigo2008}. Since the \cite{bruzual2003} models are widely used, it is worth showing a comparison between estimates of physical parameters obtained using the updated and original versions of these models.\footnote{For the purpose of this comparison, we use the `bc03.models.padova\_1994\_chabrier\_imf.tar.gz' model from the webpage http://www.bruzual.org/bc03/Updated\_version\_2012/. This includes an updated description of the fraction of stellar mass returned to the ISM by an evolving stellar population.} We thus build a CLSC\_BC03 spectral library, where the SFH and dust-attenuation priors are identical to the CLSC spectral library presented in Section~\ref{sec:basic}. We perform the comparison using this simple spectral library in order to limit the effects of other variables (e.g., the shape of the SFH and the slope of the dust attenuation law) when fitting observed SEDs.

In Fig.~\ref{fig:app:bc03}, we compare the optical and NIR colours predicted by the CLSC (orange) and CLSC\_BC03 (magenta) spectral libraries with the colours of the galaxies in the 3D-HST photometric sample. The format of this Figure is the same as that of Fig.~\ref{fig:col3lib}. The colour-colour space covered by the two spectral libraries is very similar. We observe a difference in the lower redshift bin (top panels) where the CLSC\_BC03 spectral library extends to slightly redder F435$-$F775 colours at fixed F125$-$F160 colours than the CLSC spectral library. In Fig.~\ref{fig:app:bc03param}, we compare the constraints on stellar mass (top panel) and SFR (bottom panel) derived using the CLSC and the CLSC\_BC03 spectral libraries. Symbols are colour-coded according to the difference between the estimates of SFR. The estimates of stellar mass are in agreement within a factor of two along the whole mass range. The estimates of SFR tend to be larger when derived with the CLSC\_BC03 than with the CLSC spectral library, by a median of 0.1 dex.  This discrepancy arises from the small difference in the rest-frame UV-optical colours between the two versions of the models (see the upper panels of Fig.~\ref{fig:app:bc03}).

\end{document}